\documentclass[12pt,authoryear]{elsarticle}

\usepackage{amssymb}
\usepackage{amsmath}
\usepackage{rotating}
\usepackage{adjustbox}
\usepackage{pdflscape}
\usepackage{multirow}
\usepackage{url}
\usepackage{xurl}

\usepackage{longtable}
\usepackage{tabularx}
\usepackage{rotating}

\usepackage{booktabs}

\begin{document}

\begin{frontmatter}

\title{Application of association rule mining to assess forest species distribution in Italy considering abiotic and biotic factors}

\author[label1]{Valeria Aloisi}
\ead{valeria.aloisi@unisalento.it}
\author[label2,label3]{Sergio Noce}
\ead{sergio.noce@cmcc.it}
\author[label1,label3]{Italo Epicoco\corref{cor1}}
\cortext[cor1]{Corresponding author: Italo Epicoco, e-mail address: italo.epicoco@unisalento.it}
\author[label2,label3]{Cristina Cipriano}
\ead{cristina.cipriano@cmcc.it}
\author[label1,label3]{Massimo Cafaro}
\ead{massimo.cafaro@unisalento.it}
\author[label2,label4]{Giuseppe Brundu}
\ead{gbrundu@uniss.it}
\author[label5]{Lorenzo Arcidiaco}
\ead{lorenzo.arcidiaco@cnr.it}
\author[label2,label3,label4]{Donatella Spano}
\ead{spano@uniss.it}
\author[label1,label3]{Giovanni Aloisio}
\ead{giovanni.aloisio@cmcc.it}
\author[label2,label3,label5]{Simone~Mereu}
\ead{simone.mereu@cmcc.it}

\affiliation[label1]{organization={Department of Engineering for Innovation, University of Salento},
            city={Lecce},
            country={Italy}}

\affiliation[label2]{organization={National Biodiversity Future Center (NBFC)},
            city={Palermo},
            country={Italy}}
 
\affiliation[label3]{organization={CMCC Foundation - Euro-Mediterranean Center on Climate Change},
            city={Lecce},
            country={Italy}}

\affiliation[label4]{organization={Department of agricultural sciences, University of Sassari},
            city={Sassari},
            country={Italy}}

\affiliation[label5]{organization={Institute of Bioeconomy, National Research Council of Italy (CNR)},
            city={Sesto Fiorentino (FI)},
            country={Italy}}

\begin{abstract}
Biodiversity monitoring represents a pressing global priority, and assessing forest community composition plays a crucial role due to its influence on ecosystem functions. The spatial distribution of forest species becomes essential for understanding biodiversity dynamics, territorial planning, aiding nature conservation and enhancing ecosystem resilience amid global change. Association Rule Mining, commonly applied to other scientific contexts, is now innovatively adopted in the ecological field to explore the relationships among co-occurring plant species and extract hidden interpretable patterns, also with abiotic and biotic conditions. Multiple heterogeneous data sources were integrated through data preprocessing into a unique dataset, including georeferenced information about 151 plant species monitored within 6,784 plots across Italy and several bioclimatic indices, soil-related factors, and variables from earth observations. The Frequent Pattern Growth algorithm, used for association rule mining, provided interesting and encouraging findings, suggesting ecological rules among plant species and environmental conditions. Indeed, temperature seasonality between 650-700 and precipitation seasonality between 45-50 resulted very correlated with \textit{Picea abies} (confidence = 90.9\%, lift = 7.13). Patterns detected for \textit{Picea abies} highlighted its ecological specificity, indicating a strong association with cold, highly seasonal environments, and particular plant communities. Some species appeared acting as community "hubs", frequently co-occurring with other species, suggesting ties to specific environmental or biotic conditions. These findings represent a valuable resource for future research, especially in regions with similar environmental settings and when prior ecological knowledge exists, also underlining the importance of publicly accessible, high-quality ecological data.
\end{abstract}


\end{frontmatter}

\section{Introduction}
\label{sec1}

Monitoring biodiversity has become an urgent global priority in light of escalating environmental pressures and accelerating species loss \citep{diaz2019pervasive}. The adoption of the Kunming-Montreal Global Biodiversity Framework and, at the European level, the entry into force of the Nature Restoration Regulation mark a turning point in policy ambition \citep{ma2023kunming, penca2025transformative}. These frameworks imply the need for accurate, scalable, and timely ecological information to guide restoration efforts, conservation planning, and climate adaptation strategies \citep{kissling2018building}. Forest ecosystems, which host intricate biological communities and deliver essential ecosystem services, are at the forefront of these restoration goals \citep{hua2022biodiversity,pan2011large}.

Building on current global and regional priorities, assessing forest community composition is particularly critical, as it underpins key ecosystem functions and directly influences biodiversity \citep{bertrand2011changes, brose2016biodiversity}. Understanding the spatial distribution of forest species provides essential information for biodiversity monitoring, territorial planning, and adaptive forest management, thereby supporting nature conservation and ecosystem resilience in the face of global change \citep{sullivan2017diversity, chazdon2019restoring}.

Understanding and predicting community composition has long been a central goal in ecology, giving rise to a dedicated subfield known as community ecology. This discipline focuses on the patterns of species diversity, abundance, and composition within ecological communities, and the underlying processes that shape them. While the conceptual roots of community ecology can be traced back to the pioneering work of Alexander von Humboldt and Frederic Clements in the $19^{th}$ and early $20^{th}$ centuries, it was not until the 1950s that quantitative approaches emerged, notably through the development of niche theory and competition models \citep{macarthur1967limiting}.

Species do not exist in isolation; rather, they interact through mechanisms such as competition, neutrality, and facilitation \citep{thuiller2009biomod,elith2009species}. Therefore, predicting community composition cannot rely solely on species' responses to abiotic conditions. These interactions imply that species co-occurrence is not entirely stochastic. Although a general theory of community assembly across space and time is still lacking, ecologists increasingly recognize that community composition results from both deterministic and stochastic processes, often referred to collectively as assembly processes \citep{Gravel2006399}. Four primary processes shape these dynamics \citep{Vellend2010183}: selection (differences in survival and reproduction among species), drift (random changes in how common species are), speciation (the emergence of new species), and dispersal (movement across space).
The development of Species Distribution Models (SDMs) in the late 1990s, facilitated by advances in computational capacity and the availability of Geographic Information Systems (GIS), provided tools to quantitatively link species distributions with environmental variables. SDMs, also known as ecological niche models, are now a fundamental component of community ecology. They include a variety of methods, such as Generalized Linear Models (GLMs), Generalized Additive Models (GAMs), climatic envelope models, and more recently, machine learning algorithms. SDMs are widely used not only for studying the autoecology of species, but also for predicting species responses to environmental change \citep{norberg2019comprehensive, noce2017likelihood, noce2019climate, noce2023altitudinal} and for informing conservation planning \citep{swan2021species,franklin2023species, rathore2023efficacy}.
A key limitation of SDMs in a community context is that they typically model species independently, assuming that environmental factors alone determine distributions \citep{guisan2005predicting, watling2015performance}. However, species interactions are implicitly embedded in presence data, meaning that the resulting models often reflect where species actually live under current conditions (realized niche), rather than all the places they could potentially survive if there were no competitors or other limiting factors (fundamental niche).
Joint Species Distribution Models (JSDMs) represent a major advance by modelling the distribution of multiple species simultaneously. This enables the identification of community-level patterns in species-environment relationships \citep{Ovaskainen2016428}. More sophisticated JSDMs, such as the Hierarchical Modelling of Species Communities (HMSC) framework \citep{ovaskainen2017make}, incorporate species traits and phylogenetic information, allowing improved predictions even for data-poor species. The underlying assumption is that phylogenetically related species or those sharing functional traits will respond similarly to environmental conditions.
Importantly, JSDMs estimate residual species co-occurrence, captured in a residual correlation matrix that accounts for associations unexplained by environmental covariates \citep{Pollock2014397}. While these residual correlations may suggest underlying biotic interactions, caution is warranted. As Warton et al. \citep{warton2015so} noted, such correlations can also arise from shared unmeasured environmental variables or sampling artifacts. Nonetheless, the co-occurrence matrix also offers valuable potential for conditional prediction, that is, estimating the probability of a species’ presence given the occurrence of another \citep{warton2015so}. This capacity holds promise for applied ecology, including species detection, monitoring, and management.
Most JSDMs are implemented within a Bayesian framework, which provides a coherent way to quantify uncertainty and incorporate prior knowledge. However, Bayesian models are often computationally demanding, restricting their application to small datasets, limited spatial extents, or low-resolution grids. This poses challenges for addressing urgent environmental questions that require high-resolution, large-scale predictions.

To support biodiversity conservation and restoration at scale, it is important to develop modelling frameworks that are not only ecologically comprehensive but also computationally efficient and user-friendly. Approaches that integrate multiple sources of ecological knowledge, including species interactions, environmental variables, and functional traits, can more faithfully represent community dynamics. In recent years, Artificial Intelligence (AI) and Machine Learning (ML) algorithms have been increasingly applied in terrestrial ecology to improve model performance, manage complex datasets, and uncover non-linear patterns that traditional methods may miss \citep{cipriano2025algorithms}. 

In light of this, the present study focuses on advancing community-level ecological models that are computationally efficient, transparent, and scalable. Deep Learning methods involving Neural Networks (NNs) are usually adopted for predictive tasks, but these techniques are not preferable when the goal is explainability. Indeed, representing "black-box" models, NNs offer limited interpretability and flexibility in understanding their internal structures and decision-making processes \citep{Shen_et_al_2024}. This makes it challenging to clarify the relationships between influencing factors and forest species distribution. Therefore, an approach based on Association Rule Mining (ARM) could enable a clearer interpretation due to its inherent explainability. \\
ARM is a well-established technique within the field of ML, more specifically in unsupervised learning, aimed at discovering hidden and interpretable patterns in large datasets without relying on predefined class labels or target variables \citep{han2012data}. Initially proposed for market basket analysis, ARM was successfully adopted in further research fields. Concerning the ecological context, only a small number of studies exploited ARM for forest species distribution. Indeed, few research articles emerged from a search on Scopus performed until $06^{th}$ June 2025 with the terms "association rule" and "plant species" whereas the keywords “association rule” and “forest community” did not show results yet. Similar queries and results were also obtained from the Web of Science database. Specifically, a study explored how plant trait correlations, insularity, and climate may influence plant community assembly on islands, focusing on the Southwest Pacific \citep{Ciarle_2024}. Considering 485 plant species across eight archipelagos surrounding New Zealand, the study adopted ARM to evaluate plant trait correlations, then assessed their role, together with insularity and climate, on the incidence of dioecy. Instead, concerning the assessment of plant species co-occurrence, a relevant study conducted by \citep{Silva_et_al_2016} exploited ARM techniques, based on the Apriori algorithm, to extract hidden patterns of co-occurrences among 312 forest species monitored on the Barro Colorado Island, Panama. Positive and negative correlations emerged among pairs and groups of species, thus highlighting the effectiveness of the approach in deriving knowledge, also by crossing species occurrences with habitats \citep{Silva_et_al_2016}. A similar method was then applied in \citep{Orozco-Arias_et_al_2019} to explore worldwide co-occurrences among 17 species of the genus \textit{Brachypodium}. Based on georeferenced data of the species, the Apriori algorithm for ARM revealed that \textit{Brachypodium sylvaticum} frequently co-occurs with \textit{Brachypodium pinnatum}, \textit{Brachypodium rupestre}, \textit{Brachypodium retusum}, and \textit{Brachypodium phoenicoides}, due to its wide distribution in Europe, Asia, and North of Africa \citep{Orozco-Arias_et_al_2019}. However, a limitation of the previous two studies is that they did not consider any abiotic or biotic factors that could influence the spatial distribution of the plant species. Another approach was adopted in a study identified through the literature search, which had a slightly different goal. A weighted ARM with Predictive Apriori algorithm was proposed in \citep{Pratheepa_et_al_2016} to investigate the role of some abiotic factors on the incidence of the insect pest \textit{Helicoverpaarmigera} on cotton crops. The obtained if-then rules revealed that, under specific ranges for temperature and relative humidity, the occurrence of this pest would be high \citep{Pratheepa_et_al_2016}. \cite{souza2021tree} confirmed the feasibility of applying association rule analysis to ecological datasets in tropical forests, highlighting its potential as a tool for detecting patterns of species coexistence and ecosystem functioning. A recent paper \citep{ghosh2025association} has proposed a method based on association rules to identify threatened, ecologically synergistic mangrove species in India for targeted restoration. \\
The present study aims to explore the underlying relationships that may exist among co-occurring species, also considering biotic and abiotic conditions, focusing on 151 forest species monitored in 6,784 sites across Italy, thus improving the assessment of forest community distribution. To the best of our knowledge, this is the first study to apply ARM with a more efficient algorithm, the Frequent Pattern Growth (FP-Growth), for automatically extracting hidden patterns among plant species, taking into account several climatic, soil-related, and earth observation factors.

\section{Methods}
\label{sec2}

\subsection{Data sources and database implementation}
\label{subsec2.2}

\subsubsection{Plant species data}
\label{subsec2.2.1}
The Italian National Forest Inventory (INFC), established to meet the United Nations Framework Convention on Climate Change (UNFCCC) requirements, also supports national and international forest reporting, including the Kyoto Protocol and the Food and Agriculture Organization of the United Nations (FAO) assessments. The INFC2015, in accordance with the INFC2005, follows a three-phase design with systematic sampling on a $1\times1$~Km grid \citep{gasparini2022italian}. Phase 1 involved photo-interpretation of 301,000 points using Coordination of Information on the Environment (CORINE) and FAO classifications. In Phase 2, about 30,000 forest points were field-checked for qualitative data. Phase 3 focused on 7,000 plots where quantitative data were collected. These data support statistically sound estimates of about 50 forest variables, mainly for wood and carbon assessments. We derived plot coordinates and basal area from the third phase of the INFC2015. Most of the variables used in this study were calculated within a 13-meter diameter subplot (AdS13), which serves as the core sampling area for dendrometric and structural measurements. \\ 
The species list with the associated unique identifier (UNICODE) is shown in Supplementary Table 1.

\subsubsection{Bioclimatic variables}
\label{subsec2.2.2}
Data for 20 bioclimatic variables were downloaded from two global datasets, 19 of which are from the WorldClim 2.1 database \citep{fick2017worldclim} for the 1970-2000 period and the Aridity index from Version 3 of the Global Aridity index and potential Evapotranspiration database \citep{zomer2022version}. Both datasets have a spatial resolution of 30 arcsec ($\approx 1$ Km). To avoid multicollinearity among the WorldClim variables, a pairwise correlation analysis was conducted across the 19 layers. Based on a correlation threshold of $|r| \geq 0.7$, we excluded highly correlated variables and retained only a subset for subsequent analyses: BIO1 (Annual Mean Temperature), BIO4 (Temperature Seasonality), BIO8 (Mean Temperature of Wettest Quarter), BIO12 (Annual Precipitation), and BIO15 (Precipitation Seasonality). Bioclimatic values were extracted from the above-described sets of raster layers based on the INFC2015 sampling plots, within the ESRI ArcGIS Pro ver 3.4.3 environment, using the Nearest Neighbor (NNb) technique, which assigns to each point the value of the 1-Km resolution cell in which the plot falls. 

\subsubsection{Geopedological variables}
\label{subsec2.2.3}
Physical and chemical soil properties were extracted from SoilGrids at 250 m resolution \citep{hengl2017soilgrids250m} released in May 2020 and accessed in March 2025. Variables included in the database were the volume of water content at -1500 kPa $\left(10^{-2} \ \mathrm{cm^3 \ cm^{-3}}\right) \cdot 10$) and pH Water (pH*10) for the 5-15 cm soil depth layer. 
As in the previous case, values were extracted from raster layers using the INFC2015 sampling plots within the ESRI ArcGIS Pro environment, applying the NNb technique. 
Lithological information was derived from the recent Italian lithological map provided by \cite{bucci2022new}, which offers a standardized classification of rock types across the Italian territory. Lithology data were spatially intersected with sampling plots using the same geoprocessing procedures to ensure consistency with soil variable extraction. A table describing the lithological codes and their corresponding types is provided in the Supplementary Table 2.

\subsubsection{Earth observation variables}
\label{subsec2.2.4}

\begin{figure}[h!]
\centering
\includegraphics[width=\linewidth]{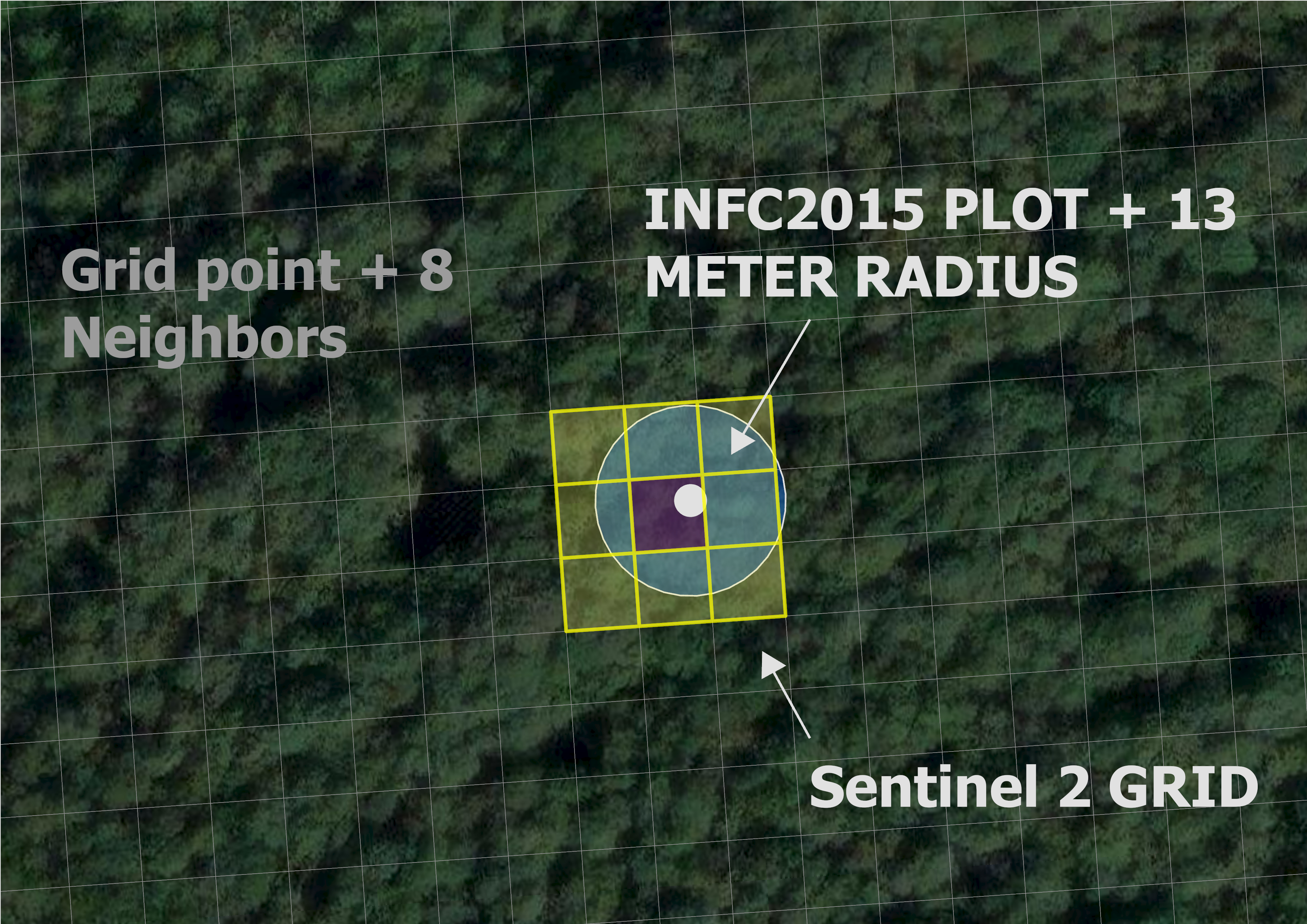}
\caption{Earth Observation data value calculation.}\label{Fig_RsAn}
\end{figure}

The quantification of spatiotemporal variations in vegetation dynamics was conducted by extracting key statistical descriptors of the Normalized Difference Vegetation Index (NDVI) using Sentinel-2 surface reflectance level 2 data and the INFC2015 sampling plots.\\
The analysis was conducted within the Google Earth Engine (GEE) platform \citep{gorelick2017google}, leveraging its scalable cloud-based geospatial processing capabilities to efficiently manage large datasets and perform high-resolution time-series computations. The methodological framework was designed to assess both intra-annual and inter-annual vegetation trends, while minimizing atmospheric noise and accounting for localized spatial variability. The INFC2015 sampling plots dataset was uploaded to GEE as a FeatureCollection, and it was used to extract spectral metrics. To capture seasonal and interannual variations in vegetation activity, the analysis was performed over a seven-year period from 2018 to 2024. For each year, the growing season, typically characterized by peak photosynthetic activity in temperate forest ecosystems, was defined as the period from March to September (both inclusive).\\
This temporal window was selected to minimize potential confounding effects from winter dormancy or snow cover in mountainous areas. For each year within the specified interval, all available Sentinel-2 Level 2A Surface Reflectance images were acquired from the Copernicus S2\textunderscore Sr\textunderscore Harmonized dataset. This harmonized collection ensures radiometric consistency across Sentinel-2A and 2B sensors, making it suitable for multi-year time-series analysis. To mitigate cloud contamination, a cloud-masking procedure was applied based on the Msk\textunderscore Cldprb band, which provides per-pixel estimates of cloud probability. For each image, pixels with cloud probabilities exceeding 20\% were excluded from analysis. This threshold was selected based on a balance between retaining sufficient data for temporal compositing and ensuring the reliability of vegetation index measurements. The cloud mask was applied before NDVI calculation to preserve only high-quality observations. This approach is particularly important in regions with frequent cloud cover, such as northern and mountainous parts of Italy. NDVI was computed for each image using the standard formula:

\begin{equation}
\text{NDVI} = \frac{\text{NIR} - \text{RED}}{\text{NIR} + \text{RED}}
\label{eq:ndvi}
\end{equation}

where the near-infrared (NIR) reflectance corresponds to band B8 and the red reflectance to band B4 of Sentinel-2 imagery. After cloud-masking, the NDVI images were stacked into yearly collections, filtered by date and spatial extent. For each year, NDVI time series were summarized using four statistical measures: minimum (\textit{NDVI\_min}), maximum (\textit{NDVI\_max}), median (\textit{NDVI\_median}), and standard deviation (\textit{NDVI\_stdDev}). These metrics capture key aspects of vegetation behavior: minimum and maximum values represent seasonal extremes, the median provides a measure of central tendency over the growing season, and the standard deviation reflects intra-seasonal variability, which may indicate land cover transitions or phenological dynamics.\\
In addition to point-based statistics, we sought to understand the local spatial variability in NDVI values surrounding each sampling point. This was achieved through a neighborhood analysis using a 3×3 pixel window (corresponding to a spatial extent of 30 m × 30 m at 10 m resolution). Specifically, focal operations were performed on the median NDVI layer to derive local maximum (vic\textunderscore max), and median (vic\textunderscore median) values.\\
Spatial variability was further quantified using a spatial standard deviation (vic\textunderscore std), computed with a 3×3 square kernel through the \textit{reduceNeighborhood GEE function}. This neighborhood-based analysis helps capture spatial heterogeneity, which can be indicative of fragmented vegetation, edge effects, or mixed land cover types within the vicinity of the point. 
For each year, NDVI summary statistics were extracted at each sampling location using the \textit{reduceRegions} function, which allows batch computation of zonal statistics across multiple points. Pixel-based NDVI values were matched with their corresponding spatial metrics using unique point identifiers. The final feature set for each year included the following attributes: Maximum NDVI 3$\times$3 and Standard Deviation of NDVI 3$\times$3, representing the highest NDVI value and spatial variability within a 3$\times$3 pixel neighborhood, respectively. For each sampling point, these values were averaged over the entire temporal series (2017-2024) to obtain a representative metric, which was then integrated into the final database (MDVIME, SDVIME). These metrics offer a spatially explicit view of vegetation conditions, supporting ecological modelling and land condition assessments by capturing local-scale vegetation dynamics influenced by environmental or anthropogenic factors.\\
Additionally, we used a global high-resolution canopy height dataset to calculate the mean vegetation height and standard deviation within each block unit of the study area. This dataset, developed by \cite{lang2023high}, integrates Sentinel-2 imagery with GEDI spaceborne LiDAR data and employs a deep learning model to estimate canopy height globally for the year 2020 at a spatial resolution of 10 meters. Given the 13-meter radius of the INFC subplot, Earth Observation indicators were calculated by averaging the values within a 3×3 cell window centered on the cell corresponding to the INFC point (Fig. \ref{Fig_RsAn}).

\subsubsection{Final database implementation}
\label{subsec2.2.6}
\begin{figure}[h!]
\centering
\includegraphics[width=\linewidth]{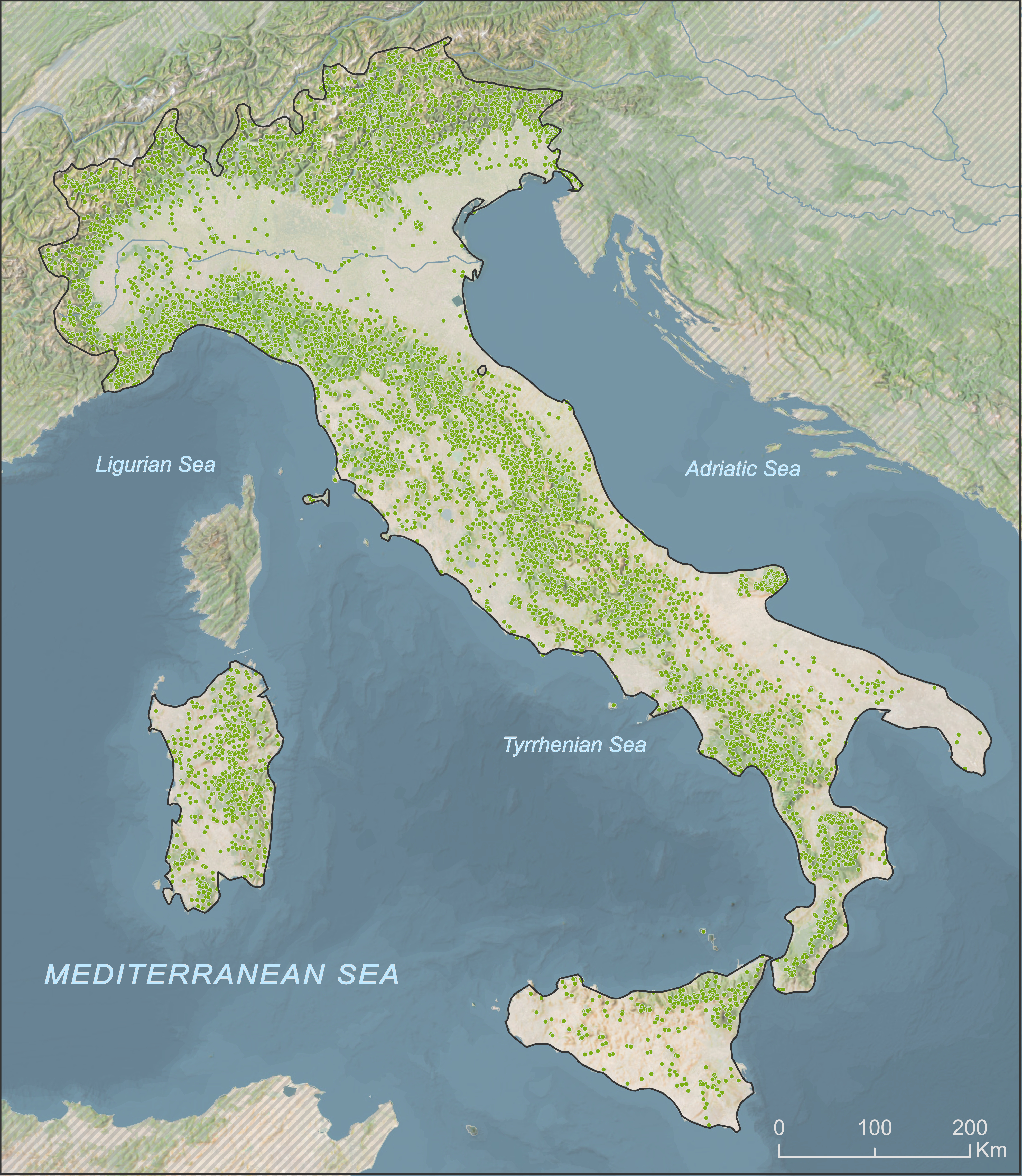}
\caption{Spatial distribution of the 6,784 monitoring sites across Italy.}\label{Fig_Italy_map}
\end{figure}

All INFC2015 plots containing one or more missing data values within the bioclimatic and soil variables were excluded. Regarding the Earth Observation data, plots with more than one missing data point within the 2017–2024 time series for both the Maximum NDVI and the Standard Deviation NDVI indices were removed (e.g., if missing data occurred in two or more years, the entire plot was excluded). From the initial set of plots belonging to INFC2015, a total of 6,784 plots were selected for the final composition of the analysis database, and they are shown in Fig. \ref{Fig_Italy_map}).

The variable naming convention adopted in this study follows a standardized schema designed to facilitate categorization and interpretation. Each variable code is composed as follows: \textit{X\_YYYYYY}, where \textit{X} is a single uppercase letter indicating the thematic category, and \textit{YYYYYY} is a unique identifier for the specific variable. For instance, the prefix \textit{P} refers to plot-level information such as plot ID, geographic coordinates, or structural attributes; \textit{S} identifies soil-related variables including physical and chemical properties; \textit{C} designates climate variables such as temperature and precipitation indices, and \textit{E} is for Earth observation variables.

This structured format ensures consistency across datasets and enhances automated data handling. As examples, the code \texttt{P\_LATITU} refers to the plot’s latitude, \texttt{S\_SGPHWA} to soil pH, and \texttt{C\_WC0001} to the annual mean temperature.
The implemented final database is described in Supplementary Table 3.

\subsection{Association rule mining}
\label{subsec2.3}

Association rule mining is one of the most important and well-researched techniques in data mining \citep{Zhao_Bhowmick_2003}, which represents the core process of the so-called Knowledge Discovery in Database (KDD). The term KDD was initially introduced in a workshop at the International Joint Conference on Artificial Intelligence in Detroit, USA, in 1989 \citep{Zhang_Wu_2011}, then it was defined in 1992 and 1996. According to the widely accepted 1996 definition, KDD represents the \textit{nontrivial process of identifying valid, novel, potentially useful, and ultimately understandable patterns in data} \citep{Fayyad_et_al_1996, Zhang_Wu_2011}. \\
Association rule mining was first proposed by \citep{Agrawal_et_al_1993} for market basket analysis \citep{Zhang_Wu_2011} to identify the sets of items that are frequently bought together at a supermarket by analyzing customer shopping carts \citep{Zaki_Meira_2020}. Then, association analysis found several applications in many other fields such as telecommunication networks, risk management, inventory control, recommendation systems, data classification and clustering, catalog design, healthcare, and loss-leader analysis \citep{Zhao_Bhowmick_2003, Kumbhare_Chobe_2014, Shen_et_al_2024, Papi_et_al_2022, Darrab_et_al_2024, Rawat_et_al_2023, Versichele_et_al_2014, Cagliero_et_al_2016}.\\
The goal of ARM is to find interesting relationships, correlations, patterns, and association structures among frequently appearing items in transactional datasets, without implying causality \citep{Zhao_Bhowmick_2003, Orozco-Arias_et_al_2019, Silva_et_al_2016, Wu_et_2008, Tan_2007}.
\\
Let $I = \{ i_1, i_2, \ldots, i_n \}$ be a set of elements called \textit{items}. For example, \textit{I} may represent the collection of all the products (e.g., milk, butter, bread, etc.) sold at a supermarket \citep{Zaki_Meira_2020}. A set $X \subseteq I$ is called \textit{itemset}. A \textit{transaction} corresponds to the set of items in an operation, such as the subset of products purchased by a particular customer in market basket analysis \citep{Silva_et_al_2016, Brin_et_al_1997}. Each transaction is usually associated with a unique identifier called \textit{Transaction IDentifier (TID)} \citep{Zhang_Wu_2011, Zaki_Meira_2020}. Given a transaction dataset \textit{D} whose records are transactions over \textit{I}, each itemset is characterized by a statistical measure called \textit{support}, denoted as \textit{supp}, indicating the number of transactions in \textit{D} that contain the itemset \citep{Zhang_Wu_2011, Zaki_Meira_2020}. An itemset \textit{X} is said to be \textit{frequent} in \textit{D} if its support is greater than or equal to a \textit{minimum support threshold}, denoted as \textit{minsup} \citep{Zhang_Wu_2011, Zaki_Meira_2020}. If this threshold is specified as a fraction by users or experts, it must be referred to the so-called \textit{relative support} of an itemset, indicating the fraction of transactions in \textit{D} that contain the itemset \citep{Zaki_Meira_2020}. Given these premises, an association rule is an implication in the form of $X \rightarrow Y$, where \textit{X} and \textit{Y} are disjoint itemsets, that is $X, Y \subseteq I$ and $X \cap Y = \varnothing$ \citep{Zhao_Bhowmick_2003, Zhang_Wu_2011, Zaki_Meira_2020}. In this case, \textit{X} is called \textit{antecedent}, whereas \textit{Y} is named \textit{consequent} \citep{Zhao_Bhowmick_2003}. Therefore, an association rule detects a relationship between the itemsets \textit{X} and \textit{Y}, thus providing information in the form of if-then statements \citep{Zhang_Wu_2011}. For example, in the context of market basket analysis, association rules may detect that ‘bread’ and ‘butter’ are frequently brought together, suggesting that if a customer buys bread, then they may also purchase butter \citep{Kumbhare_Chobe_2014}. There are three key metrics for evaluating the quality of the association rules: \textit{support}, \textit{confidence}, and \textit{lift}. Specifically, given an association rule $X \rightarrow Y$ with $X, Y \subseteq I$ and $X \cap Y = \varnothing$, the \textit{support} of the rule is the number of transactions in which both \textit{X} and \textit{Y} co-occur as subsets \citep{Zaki_Meira_2020}:

\begin{equation}
supp(X \rightarrow Y) = supp (X \cup Y)
\end{equation}
\newline
Thus, the \textit{relative support} of the rule, denoted as \textit{rsupp}, is the fraction of transactions containing $X \cup Y$ with respect to the total number of transactions in \textit{D} \citep{Zaki_Meira_2020}:

\begin{equation}
rsupp(X \rightarrow Y) = \frac{supp (X \cup Y)}{total\ number\ of\ transactions}
\end{equation}
\newline
The \textit{confidence} of the rule is defined as the conditional probability that a transaction contains \textit{Y} given that it contains \textit{X} \citep{Zaki_Meira_2020}. Hence, it can be computed by dividing the number of transactions that contain $X \cup Y$ by the total number of transactions that contain X \citep{Zhao_Bhowmick_2003}:

\begin{equation}
conf(X \rightarrow Y) = P(Y| X)= \frac{P(X \cap Y)}{P(X)} = \frac{supp (X \cup Y)}{supp(X)}
\end{equation}
\newline
Therefore, the confidence of an association rule provides information about the strength of that rule; for example, if a rule has a confidence of 80\%, it means that 80\% of the transactions that contain \textit{X} also contain \textit{Y} together \citep{Zhao_Bhowmick_2003}. \\
The \textit{lift}, also known as \textit{interest}, of an association rule can be considered a measure of its importance \citep{Silva_et_al_2016}, and it is defined as follows \citep{Zaki_Meira_2020}:

\begin{equation}
lift(X \rightarrow Y) = \frac{P(X \cup Y)}{P(X) * P(Y)} = \frac{rsupp (X \cup Y)}{rsupp(X)*rsupp(Y)} = \frac{conf(X \rightarrow Y)}{rsupp(Y)}
\end{equation}
\newline
Specifically, a lift of 1 reveals that the antecedent and the consequent of the corresponding rule are independent \citep{Silva_et_al_2016}, whereas if the lift is greater than or less than one, this suggests that the antecedent and the consequent itemsets have a positive or a negative correlation, respectively \citep{Silva_et_al_2016}. Moreover, the lift metric provides a deeper insight into the relationships between association rules by quantifying whether the predictive power of a rule exceeds what would be expected by chance \citep{Shen_et_al_2024}. Indeed, it allows distinguishing the significant associations (high lift values) from those due to random variations in the data (low lift values).\\
ARM consists in extracting those association rules that satisfy predefined minimum support and confidence at the same time \citep{Zhang_Wu_2011, Zhao_Bhowmick_2003, Agrawal_Srikant_1994, Kumbhare_Chobe_2014}. This condition is usually known as the \textit{supp-conf framework} \citep{Zhang_Wu_2011, Agrawal_et_al_1993}. Consequently, the ARM problem is usually decomposed into two subproblems \citep{Zhao_Bhowmick_2003}: \textit{(i)} detecting the itemsets with a support greater than, or equal to, the predefined minimum support threshold (i.e., identifying all the frequent itemsets) \citep{Zhang_Wu_2011}; \textit{(ii)} given the frequent itemsets, generating those association rules that have a confidence exceeding the minimum confidence threshold \citep{Zhang_Wu_2011, Zhao_Bhowmick_2003}. In this way, the ARM becomes a two-step process \citep{Zhang_Wu_2011}. The second sub-problem is straightforward once the first one is solved \citep{Zhao_Bhowmick_2003}, hence several approaches were developed in recent years to address it. Starting from the first algorithm, AIS (from Agrawal, Imielinski, and Swami who proposed it \citep{Agrawal_et_al_1993}), the Apriori algorithm was conceived \citep{Agrawal_Srikant_1994}. It is based on a level-wise search for mining the frequent itemsets, where those containing \textit{k} items are exploited to explore those including \textit{k+1} items \citep{Kumbhare_Chobe_2014}. This searching mechanism is known as the \textit{candidate generation process} \citep{Kumbhare_Chobe_2014}. Although the Apriori algorithm represented an improvement with respect to previous approaches, it had two main drawbacks: \textit{(i)} the complexity of the candidate generation process which consumes a lot of resources in terms of time, space, and memory; and \textit{(ii)} the fact that it requires multiple scans of the dataset \citep{Zhao_Bhowmick_2003, Kumbhare_Chobe_2014}. To overcome these bottlenecks, tree-structured approaches were investigated \citep{Zhao_Bhowmick_2003}. The FP-Growth algorithm was first proposed by \citep{Han_Pei_2000}. It is based on the construction of the Frequent Pattern Tree (FP-Tree) \citep{Han_et_al_2000}, a compressed and more efficient representation of the dataset. Hence, the procedure is composed of two steps: the construction of the FP-Tree and the generation of the frequent patterns from the FP-tree \citep{Kumbhare_Chobe_2014}. This algorithm requires only two scans over the dataset for creating the FP-Tree. The first scan is used to compute the F-List, the list of frequent items sorted by frequency in descending order, whereas the second pass is needed to compress the dataset in the FP-Tree \citep{Kumbhare_Chobe_2014}. Consequently, the problem of extracting the most frequent itemsets is converted to searching and constructing trees recursively \citep{Kumbhare_Chobe_2014} without any candidate generation process. This way, FP-Growth proved to be more efficient than the previous algorithms, being an order of magnitude faster than Apriori \citep{Zhao_Bhowmick_2003, Kumbhare_Chobe_2014}.

\subsection{Data preparation}
\label{subsec2.4}

As described in Section \ref{subsec2.2}, several heterogeneous data sources related to various factors that may influence the forest species distribution were integrated through data preprocessing steps into a unified dataset. This dataset included information about the presence or absence of 151 plant species within 6,784 monitoring plots across Italy. Each plot was characterized by the geographical coordinates (i.e., latitude and longitude) and a unique identifier, named \textit{ID plot}. More than one species could be detected in a specific plot. To apply the ARM procedure, the dataset was transformed to achieve the transactional format, where each record represents a transaction. Specifically, in this case, the set \textit{I} of all the items corresponds to the set of all the 151 plant species and a transaction over \textit{I} can be thought as a subset of plant species detected in a single plot. Therefore, the transaction dataset presented a record for each plot tracking the list of those species found in that plot, as depicted in Table \ref{table_Transactions_dataset}.

\begin{table}[b]
\caption{Example with three transactions and eight species.}\label{table_Transactions_dataset}

\centering
\begin{tabular}{l l}
  \hline
  ID plot & Items (plant species) \\
  \hline
  1 & LARDEC, PICABI, PINCEM, PINSYL \\
  2 & ALNINC, FRAEXC \\
  3 & BETPEN, PICABI, POPTRE \\
  \hline
\end{tabular}
\end{table}

Moreover, several abiotic and biotic factors were extracted, preprocessed, and associated with each plot based on the geographical coordinates. The abiotic variables included five climatic drivers (i.e., annual mean temperature, temperature seasonality, mean temperature of the wettest quarter, annual precipitation, and precipitation seasonality) and four soil-related factors: aridity index, lithology, soil water content (SGWC33), and soil pH (SGPHWA). Instead, the following features derived from Earth observations were considered as biotic components: ETH Global Height, NDVI Max (temporally averaged over 2017-2024), and NDVI Std Dev (temporally averaged over 2018-2024).\\
Due to the continuous nature of these factors, a categorization procedure was needed to proceed with the association analysis. The criterion used to categorize each variable is reported in Table \ref{table_Categorizations}. Then, a one-hot encoding step was performed. Consequently, the final dataset consisted of 6,784 records corresponding to plots and 433 boolean columns, of which 151 indicated the presence or absence of each species, whereas the remaining 282 were derived from the categorization of the abiotic and biotic features. Therefore, in this dataset, the transaction described by a record indicates those plant species detected in a plot, together with the abiotic and biotic conditions associated with that plot. \\
The ARM procedure was performed on this dataset by using the Python package \textit{mlxtend}, and the minimum support threshold set in the FP-Growth algorithm was equal to 0.01. Setting a higher threshold could restrict the analysis to only the predominantly common species \citep{Silva_et_al_2016}, potentially overlooking ecologically relevant but less common patterns. Conversely, the chosen minimum support value enabled the identification of a higher number of most frequent itemsets from which to extract rarer and more interesting patterns. Moreover, a threshold of 0.07 was set on the confidence metric to filter the resulting association rules, thus ensuring a high level of precision and reliability.

\begin{table}[t]
\caption{Categorization of the continuous abiotic and biotic factors.}\label{table_Categorizations}

\centering
\adjustbox{max width=\textwidth}{
\begin{tabular}{l l l}
  \hline
  Variable & Code & Categorization \\
  \hline
  \textit{Abiotic factors}\\
  \hline

  Annual mean temperature (°C) & Bio1 & 40 classes for each 0.5 °C change \\
  
  Temperature seasonality (standard deviation ×100) & Bio4 & 7 classes for each 50 units \\
  
  Mean temperature of the wettest quarter (°C) & Bio8 & 47 classes for each 0.5 °C change \\
  
  Annual precipitation (mm) & Bio12 & 52 classes for each 50 mm change \\
  
  Precipitation seasonality (Coefficient of Variation) & Bio15 & 14 classes for each 5 units \\
  
  Aridity index & ARIIND &20 classes in 5\% percentiles \\
  
  Soil WC33 (10$^{-2}$ x cm$^{3}$ x cm$^{-3}$) x 10 & SGWC33 & 20 classes at evenly spaced intervals \\
  
  Soil PHWA (pH x 10) & SGPHWA & 5 classes at evenly spaced intervals \\
  
  Lithology (already categorical) &  & - \\
  \hline
  \textit{Biotic factors} \\
  \hline
  ETH Global Height & GLH & 21 classes for each 2 m change \\
  
  NDVI Max (temporally averaged over 2017-2024) & MDVIME & 20 classes in 5\% percentiles \\
  
  NDVI Std Dev (temporally averaged over 2018-2024) & SDVIME & 20 classes in 5\% percentiles \\
  
  \hline
\end{tabular}
}
\end{table}

\section{Results}
\label{sec3}

\begin{figure}[h!]
\centering
\includegraphics[width=\linewidth, height=0.85\textheight]{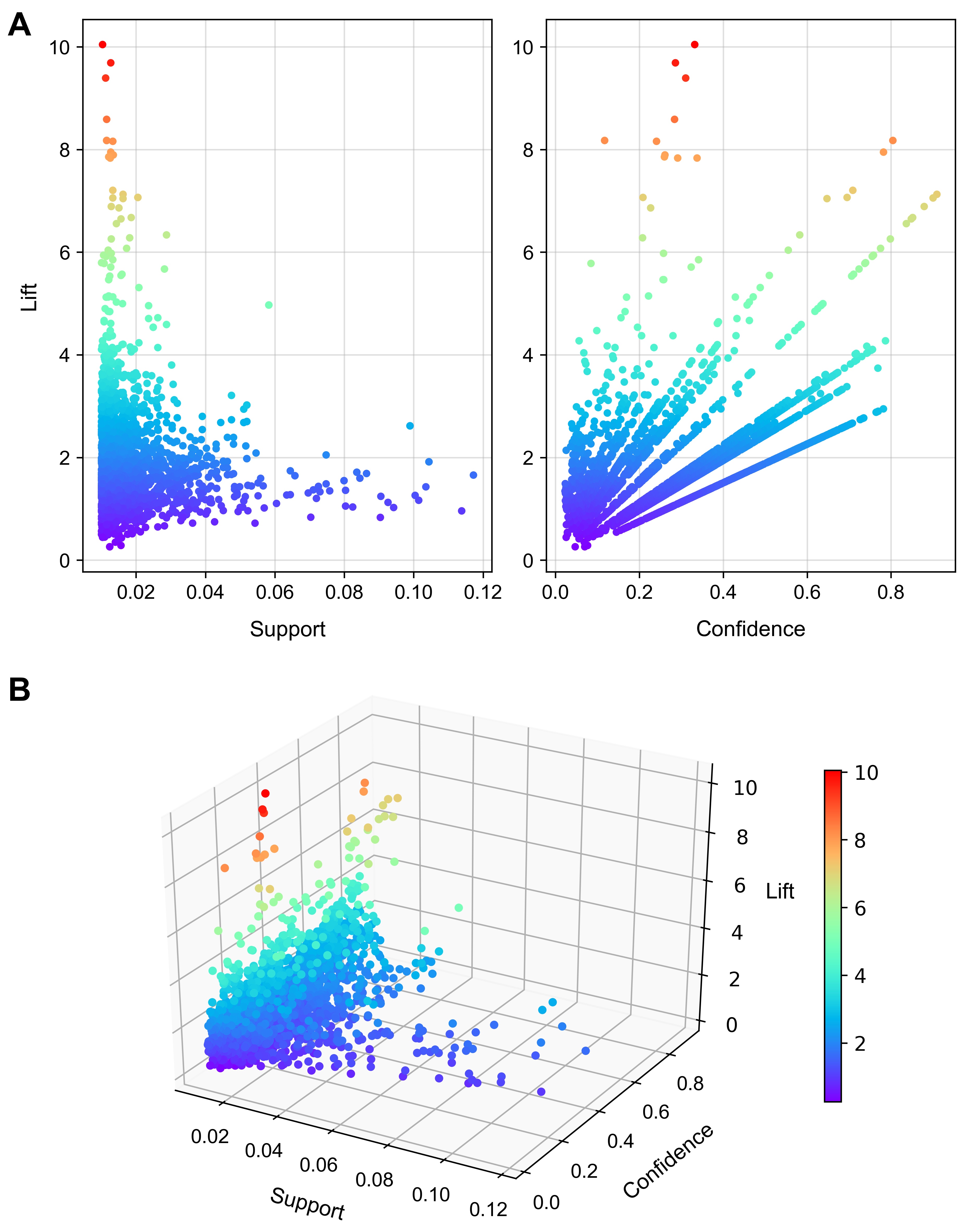}
\caption{Distribution of 2,783 association rules based on support, confidence, and lift. Panel A provides bidimensional visualizations of the association rules, with the support or confidence values on the x-axis and the lift on the y-axis. Panel B displays the association rules according to the three metrics simultaneously in a three-dimensional scatterplot.}\label{Fig_Scatterplots}
\end{figure}

The FP-Growth algorithm was applied to the transaction dataset and identified 3,925 most frequent itemsets with a relative support greater than or equal to 0.01. These itemsets included plant species, abiotic, and biotic categorical classes as items. Then, the ARM procedure produced a total of 15,548 association rules and 2,783 out of them had consequent itemsets including only plant species. These resulting 2,783 association rules are shown in Fig. \ref{Fig_Scatterplots} with respect to the support, confidence, and lift metrics. Specifically, Panel A of Fig. \ref{Fig_Scatterplots} presents two bidimensional scatterplots where each point represents a rule plotted according to its support or confidence value on the x-axis, and its lift on the y-axis. A comprehensive visualization of the 2,783 association rules according to the three metrics at once is provided in Panel B of Fig. \ref{Fig_Scatterplots} through a three-dimensional scatterplot, where the rules are also colored by lift. As emerges from this figure, higher support values correspond to less interesting association rules (i.e., low lift values), whereas support values close to the threshold result in non-trivial and more interesting association patterns (i.e., very high lift values).

Not all 2,783 association rules are strong enough and relevant, hence a confidence threshold equal to 0.7 was set to assure high precision and reliability. Thus, 44 association rules with a confidence greater than or equal to 0.7 resulted and are reported in Table \ref{table_Results_44_Assoc_rules}.

\begin{table}[h!]
\caption{Results. Association rules with support $ \geq 0.01$ and confidence $ \geq 0.7$, sorted in descending order by lift values.}\label{table_Results_44_Assoc_rules}

\centering
\adjustbox{max width=\textwidth}{
\begin{tabular}{l l l l l l}
  \hline
  No & Antecedent & Consequent & Support & Confidence & Lift \\
  \hline

1	& PHILAT & QUEILE &	0.011 & 0.804 & 8.179 \\
2	& ARBUNE, bio4 (550.0-600.0) & QUEILE &	0.013 &	0.782 &	7.952 \\
3	& ARBUNE, SGPHWA (61.2-67.8) & QUEILE &	0.013 &	0.709 &	7.208 \\
4	& bio4 (650.0-700.0), bio15 (45.0-50.0) & PICABI & 0.016 & 0.909 & 7.130 \\
5	& SGPHWA (48.0-54.6), LARDEC & PICABI &	0.013 & 0.900 &	7.058 \\
6	& ABIALB, bio4 (650.0-700.0) & PICABI &	0.013 &	0.879 &	6.892 \\
7	& bio4 (650.0-700.0), LARDEC & PICABI &	0.019 &	0.851 &	6.677 \\
8	& LARDEC, bio15 (40.0-45.0) & PICABI &	0.016 &	0.848 &	6.651 \\
9	& bio15 (45.0-50.0), LARDEC & PICABI &	0.014 &	0.836 &	6.558 \\
10	& bio4 (600.0-650.0), SGPHWA (48.0-54.6) & PICABI &	0.013 &	0.798 &	6.260 \\
11	& bio4 (650.0-700.0), bio15 (40.0-45.0) & PICABI &	0.017 &	0.775 &	6.077 \\
12	& bio4 (650.0-700.0), SGPHWA (48.0-54.6) & PICABI &	0.011 &	0.758 &	5.944 \\
13	& bio1 (5.0-5.5)°C & PICABI &	0.011 &	0.755 &	5.921 \\
14	& bio1 (4.5-5.0)°C & PICABI &	0.010 &	0.739 &	5.797 \\
15	& FAGSYL, LARDEC & PICABI &	0.011 &	0.737 &	5.783 \\
16	& SGPHWA (48.0-54.6) & PICABI &	0.028 &	0.723 &	5.674 \\
17	& Lithology class  12, LARDEC & PICABI &	0.016 &	0.711 &	5.572 \\
18	& GLH (32.0-34.0) & PICABI &	0.012 &	0.706 &	5.536 \\
19	& QUEPUB, ACEOPA & OSTCAR &	0.010 &	0.787 &	4.275 \\
20	& bio1 (7.5-8.0)°C, bio4 (600.0-650.0) & FAGSYL &	0.010 &	0.742 &	4.170 \\
21	& FRAORN, bio4 (700.0-750.0), SGPHWA (54.6-61.2) & OSTCAR &	0.010 &	0.756 &	4.107 \\
22	& FRAORN, bio4 (700.0-750.0), Lithology class  12 & OSTCAR &	0.012 &	0.752 &	4.089 \\
23	& FRAORN, ACEOPA & OSTCAR &	0.012 &	0.743 &	4.041 \\
24	& FRAORN, bio1 (10.5-11.0)°C & OSTCAR &	0.014 &	0.742 &	4.033 \\
25	& Lithology class  12, bio1 (8.0-8.5)°C & FAGSYL &	0.011 &	0.717 &	4.030 \\
26	& bio4 (600.0-650.0), bio1 (8.0-8.5)°C & FAGSYL &	0.013 &	0.714 &	4.015 \\
27	& FRAORN, SGPHWA (54.6-61.2), Lithology class  12 & OSTCAR &	0.011 &	0.730 &	3.968 \\
28	& bio4 (650.0-700.0), ACEOPA & OSTCAR &	0.012 &	0.718 &	3.903 \\
29	& FRAORN, ACEPSE & OSTCAR &	0.013 &	0.713 &	3.876 \\
30	& FRAORN, FAGSYL & OSTCAR &	0.013 &	0.703 &	3.822 \\
31	& bio4 (700.0-750.0), OSTCAR, QUEPUB & FRAORN &	0.011 &	0.768 &	3.742 \\
32	& GLH (10.0-12.0), bio4 (650.0-700.0) & QUEPUB &	0.010 &	0.782 &	2.949 \\
33	& bio15 (20.0-25.0), OSTCAR, bio4 (650.0-700.0), SGPHWA (67.8-74.4) & QUEPUB &	0.010 &	0.769 &	2.902 \\
34	& FRAORN, OSTCAR, bio4 (650.0-700.0), SGPHWA (67.8-74.4) & QUEPUB &	0.011 &	0.765 &	2.888 \\
35	& FRAORN, bio15 (20.0-25.0), OSTCAR, bio4 (650.0-700.0) & QUEPUB &	0.012 &	0.764 &	2.883 \\
36	& GLH (12.0-14.0), bio4 (650.0-700.0), SGPHWA (67.8-74.4) & QUEPUB &	0.011 &	0.735 &	2.772 \\
37	& FRAORN, bio15 (20.0-25.0), bio4 (650.0-700.0), SGPHWA (67.8-74.4) & QUEPUB &	0.010 &	0.732 &	2.762 \\
38	& FRAORN, OSTCAR, SGPHWA (67.8-74.4) & QUEPUB &	0.019 &	0.726 &	2.740 \\
39	& FRAORN, GLH (12.0-14.0) & QUEPUB &	0.014 &	0.708 &	2.670 \\
40	& FRAORN, SGWC33 (355.1-363.8), bio4 (650.0-700.0) & QUEPUB &	0.011 &	0.705 &	2.659 \\
41	& ACEMON & QUEPUB &	0.010 &	0.704 &	2.657 \\
42	& bio15 (20.0-25.0), OSTCAR, SGPHWA (67.8-74.4) & QUEPUB &	0.013 &	0.704 &	2.656 \\
43	& FRAORN, bio15 (20.0-25.0), bio4 (650.0-700.0) & QUEPUB &	0.020 &	0.704 &	2.655 \\
44	& FRAORN, bio15 (20.0-25.0), SGPHWA (67.8-74.4) & QUEPUB &	0.016 &	0.701 &	2.644 \\

  \hline
\end{tabular}
}
\end{table}

\begin{figure}[h!]
\centering
\includegraphics[width=\linewidth]{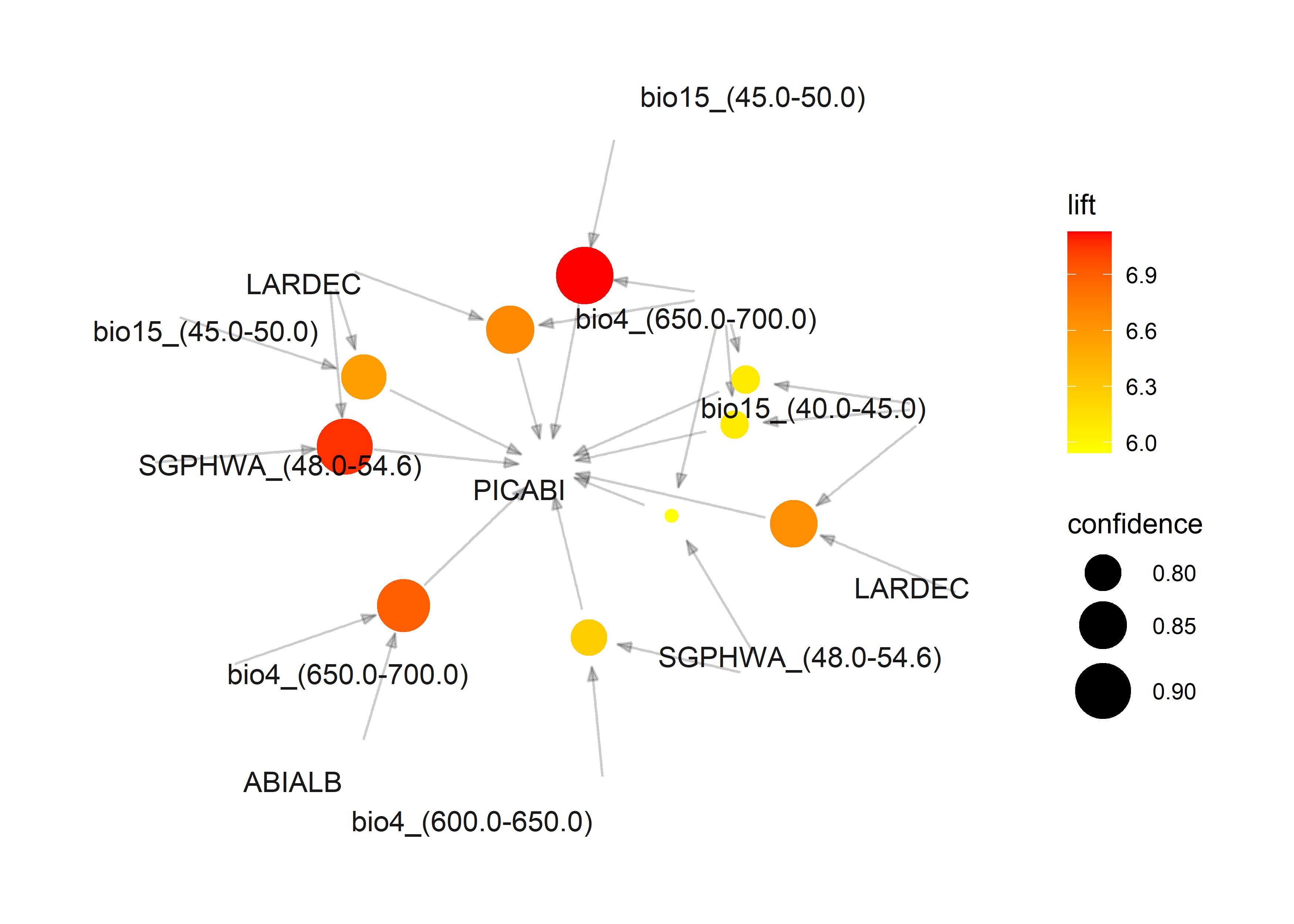}
\caption{Graph-based visualization of association rules with confidence $ \geq 0.7$ presenting \textit{Picea abies} (PICABI) as the consequent species.}\label{Fig_PICABI}

\end{figure}

\begin{figure}[h!]
\centering
\includegraphics[width=\linewidth]{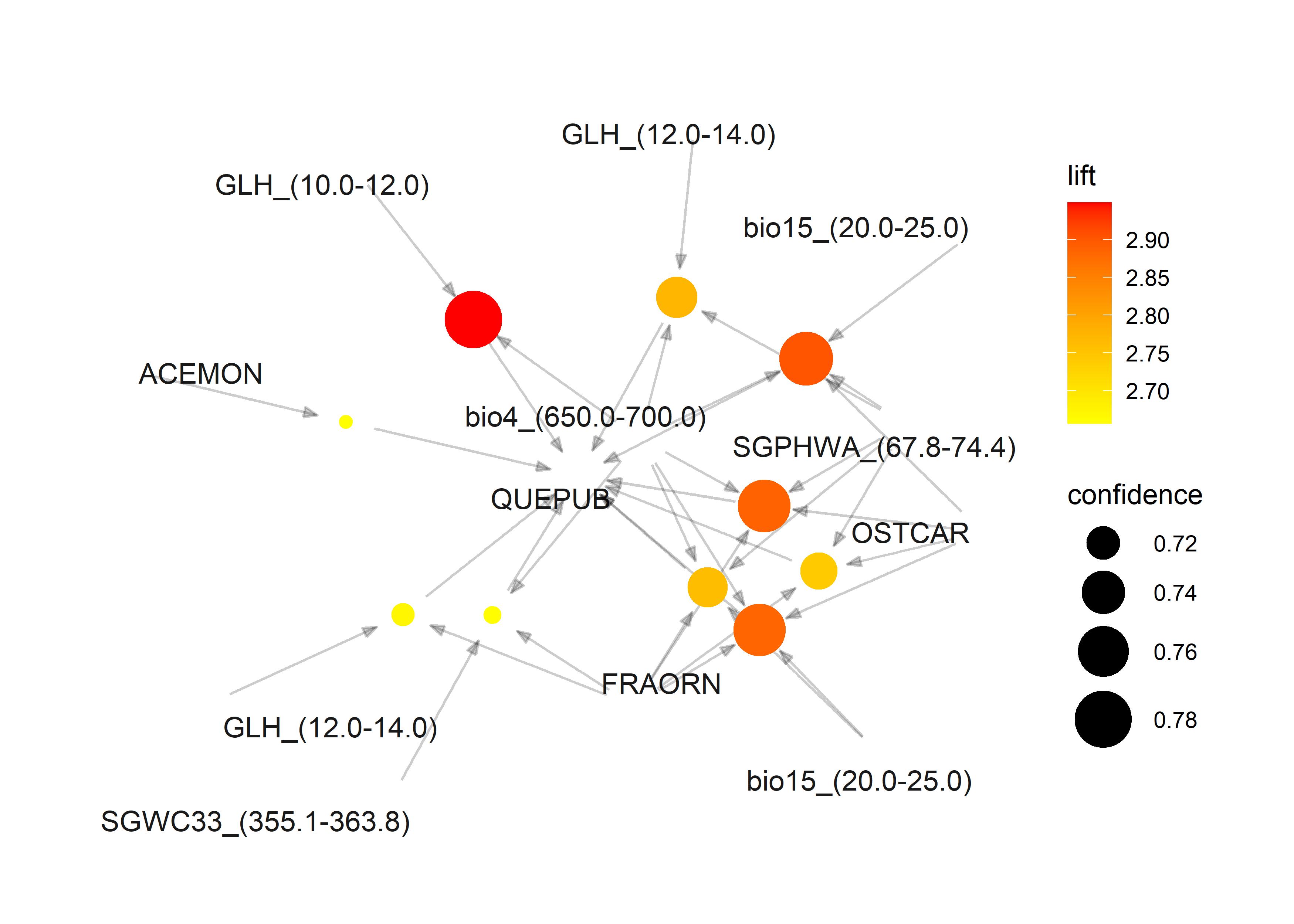}
\caption{Graph-based visualization of association rules with confidence $ \geq 0.7$ presenting \textit{Quercus pubescens} (QUEPUB) as the consequent species.}\label{Fig_QUEPUB}

\end{figure}

\begin{figure}[h!]
\centering
\includegraphics[width=\linewidth]{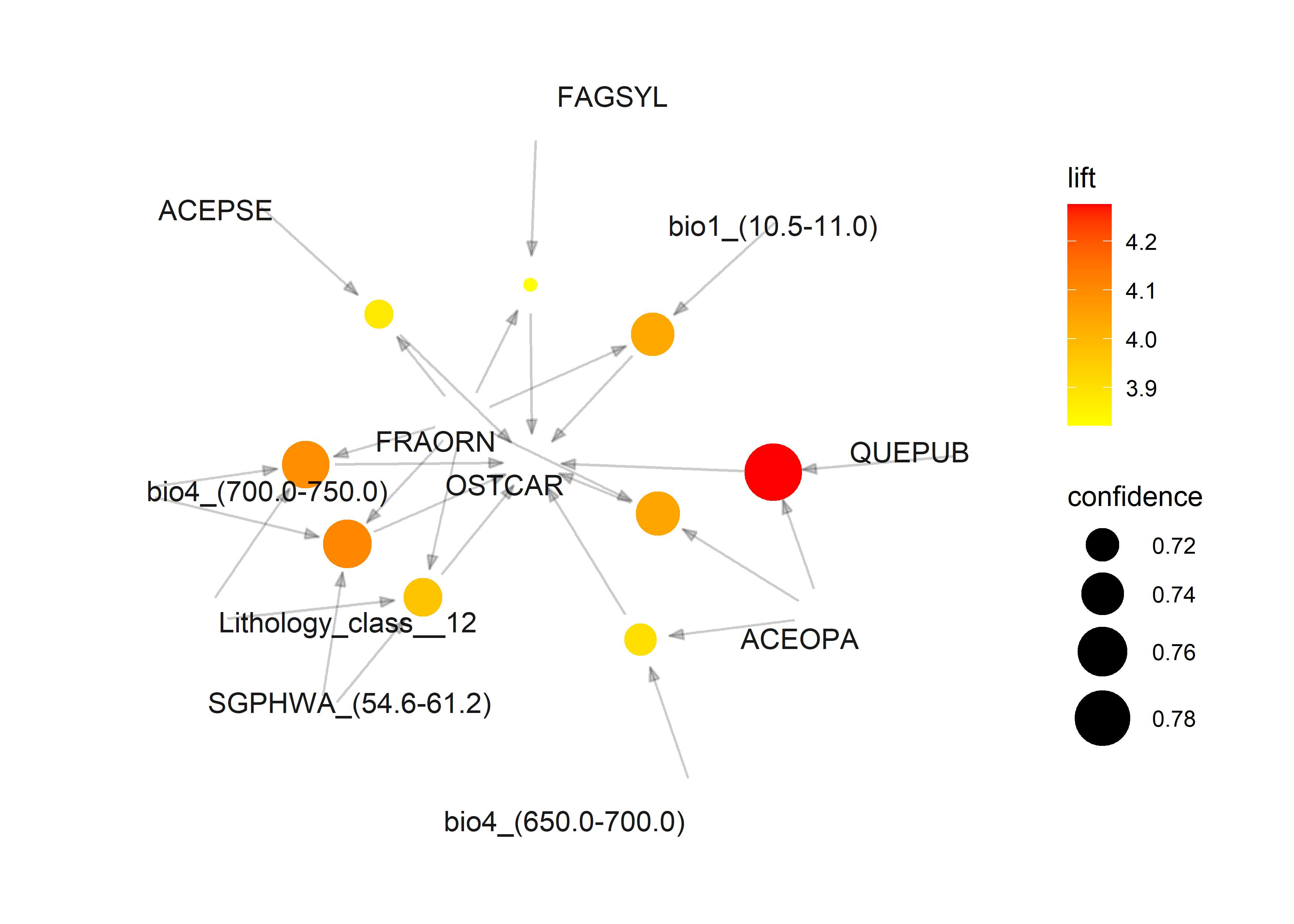}
\caption{Graph-based visualization of association rules with confidence $ \geq 0.7$ presenting \textit{Ostrya carpinifolia} (OSTCAR) as the consequent species.}\label{Fig_OSTCAR}

\end{figure}

For example, the first association rule $PHILAT \rightarrow QUIELE$ highlights that 80.4\% of the transactions that contain \textit{Phillyrea latifolia} also contain \textit{Quercus ilex} together, thus suggesting an interesting co-occurrence pattern. Moreover, it is noteworthy that the antecedent itemsets of these rules often include not only plant species but also abiotic and biotic categorical factors. Consequently, these association rules provide information concerning the abiotic and biotic conditions under which specific plant species are likely to be present with a certain level of confidence, both individually and in association with other ones. For instance, rule no.3 suggests there exists a probability of nearly 71\% that, in the presence of \textit{Arbutus unedo} (\textit{ARBUNE}) and with a soil condition in terms of pH water between 61.2 and 67.8, also \textit{Quercus ilex} (\textit{QUEILE}) will be present. \\
Additionally, several rules indicate very strong relationships between environmental variables and specific species. For example, Rule no.4, with a confidence of 90.9 \% and a lift of 7.13, reveals a highly significant ecological pattern: the combination of Bio4 (temperature seasonality between 650.0 – 700.0) and Bio15 (precipitation seasonality between 45.0 – 50.0) is very correlated with the presence of \textit{Picea abies} (\textit{PICABI}). More broadly, \textit{PICABI} appears as a consequent in 15 different rules, consistently showing high confidence values (ranging from 72.3\% to 90.9\%) and lift values above 5.5, confirming strong associations with both climatic variables (e.g., Bio1, Bio4, Bio15) and co-occurring taxa such as \textit{Larix decidua }(\textit{LARDEC}) and soil properties SGPHWA. In contrast, association rules involving \textit{Quercus pubescens} (\textit{QUEPUB}) show moderately strong patterns, with lower lift values (mostly between 2.6 and 2.9), although still above the minimum threshold of interest. These rules tend to include more diverse and complex antecedent itemsets, often involving multiple bioclimatic variables (e.g., Bio4, Bio15), soil properties (e.g., SGPHWA, GLH), and other species such as \textit{Fraxinus ornus} (\textit{FRAORN}) and \textit{Ostrya carpinifolia} (\textit{OSTCAR}). \\
Focusing on the lift metric, all association rules in Table \ref{table_Results_44_Assoc_rules} resulted to be important, showing lift values greater than 1.2, which is generally recognized as a threshold for identifying interesting rules. Specifically, their lift ranges from 2.6 to 8.2, thus highlighting those rules that are deemed valuable. \\
Figures \ref{Fig_PICABI}-\ref{Fig_OSTCAR} provide a graph-based visualization of association rules which is especially useful when exploring moderately-sized rule sets, as it highlights the structure and strength of relationships between items. Specifically, Figure \ref{Fig_PICABI} depicts the association rules where \textit{Picea abies}(\textit{PICABI}) is the consequent species (rules no. 4-18 in Table \ref{table_Results_44_Assoc_rules}), whereas Figures \ref{Fig_QUEPUB} and \ref{Fig_OSTCAR} show those that have \textit{Quercus pubescens} (\textit{QUEPUB}) and \textit{Ostrya carpinifolia} (\textit{OSTCAR}) as the consequent item, respectively. In these figures, each association rule is represented in the graph as a round node, while the plain text nodes correspond to the items. Therefore, directed edges (arrows) show the flow from antecedent items to the rule node and from the rule node to the consequent item, which is located at the center of the graph. This structure makes it easy to distinguish which items are on the left-hand side (LHS) and right-hand side (RHS) of each rule. Moreover, the size and the color of a rule node indicate the confidence and the lift of that rule, respectively.

\section{Discussion}
\label{sec4}

The present study proposed an innovative application of a traditional data mining technique, usually employed in other scientific contexts, to the ecological and biodiversity monitoring fields for assessing forest community distribution. Indeed, association rule mining was adopted to uncover interpretable patterns among plant species, abiotic conditions, biotic factors, and co-occurrence dynamics, providing insights into the ecological “rules” governing species presence and interactions. \\
Although association rules should not be considered as causal implications due to their inherent probabilistic nature, the proposed approach provided interesting and encouraging findings, suggesting interpretable "if-then" rules among plant species, biotic and abiotic drivers. \\
Despite the large number of species present in the dataset, only a limited subset appears as consequents in the association rules with confidence values exceeding the threshold, as if the rules were polarized around these species. Ecologically, this suggests that certain species may act as key “hub” species within the community, exhibiting strong and frequent co-occurrence patterns with multiple antecedents. These species might represent dominant or ecologically influential taxa whose presence is strongly linked to particular environmental conditions or biotic interactions. \\
Another relevant pattern emerging from the results is the presence of ecologically coherent species groups. The association rules mined highlight a clear tendency for some broadleaf species to co-occur in a structured and ecologically consistent way. Notably, \textit{Quercus pubescens} (QUEPUB) and \textit{Acer opalus} (ACEOPA) are often associated with \textit{Ostrya carpinifolia} (OSTCAR), as shown by the rule QUEPUB, ACEOPA → OSTCAR, suggesting a meaningful ecological relationship between these species. Similarly, \textit{Fraxinus ornus} (FRAORN) and ACEOPA also tend to co-occur with OSTCAR, as indicated by the rule FRAORN, ACEOPA → OSTCAR. Other relevant combinations, such as FRAORN with specific climatic conditions and lithological settings (carbonate rocks), reinforce the central role of OSTCAR in these associations. Altogether, these findings point to a recurring assemblage of species sharing both ecological preferences and spatial distribution patterns, likely reflecting the structure of temperate deciduous forests in the study area. The patterns identified for \textit{Picea abies} (\textit{PICABI}) emphasize the ecological specificity of this species, which seems closely tied to cold and marked seasonal climatic variation, as well as particular plant communities.
The comparatively lower lift may suggest that \textit{Quercus pubescens} (\textit{QUEPUB}) is more ecologically generalist or able to thrive across a wider range of conditions. \\
The present study has several strengths. First, it adopts a more computationally efficient algorithm, FP-Growth, for ARM among plant species compared to previous studies \citep{Zhao_Bhowmick_2003, Kumbhare_Chobe_2014}. Consequently, hidden patterns are automatically extracted in a few seconds from a large amount of ecological data. This rapid processing capability is particularly valuable when dealing with ecosystems or forest communities for which limited prior ecological knowledge is available, allowing the generation of preliminary hypotheses and guiding further ecological investigation. This makes the proposed approach a valid alternative, in terms of memory and processing time, to more traditional modelling techniques such as species distribution models (SDMs) and joint species distribution models (JSDMs), which are typically used for analyzing individual or community-level species responses to environmental gradients. \\
Unlike SDM or other approaches, which require model specification, assumptions about species-environment relationships, and often complex fitting procedures, the ARM methodology enables the rapid identification of co-occurrence patterns or associations with ecological drivers, without requiring strong assumptions. However, it is essential to emphasize that this approach is not intended as a replacement for these methods, but rather as a complementary tool in an integrated framework. If used alongside SDMs and JSDMs, ARM could provide useful insights into multispecies interactions and assemblage patterns that may be overlooked in species-by-species analyses. \\
Second, several environmental and earth observation variables were considered in the mining process as potential abiotic and biotic drivers, allowing the identification of ecological conditions that may favor the presence of a specific plant species assemblage. This contributes to a better understanding of species-environment relationships and can support the extrapolation of ecological patterns to other regions with similar environmental settings, thus helping the design of biodiversity conservation and management strategies. This aspect is especially useful when prior knowledge about the presence of certain species in a similar area is available, hence making it possible to hypothesize other species that are likely to be present as well. 
Furthermore, the interpretability of association rules and their potential to be transferred across different ecological contexts represent a unique strength. \\
However, there are also some limitations. Indeed, the present study does not provide causality patterns, but probabilistic association findings strictly related to the transaction dataset on which the ARM procedure is performed. This is due to the fact that the key metrics (i.e., support, confidence, and lift) adopted to evaluate and filter the association rules are by definition conditioned on the number and type of plant species transactions detected across Italy. Consequently, the strength and the relevance of a resulting association rule should not be interpreted in absolute terms, but it is inherently linked to the working transaction dataset. Therefore, this clearly highlights the crucial importance of developing increasingly comprehensive ecological datasets, including large volumes of data with high variability in plant species, to be as representative as possible of the various existing ecosystems and thereby improve the generalizability of the results. Moreover, the proposed approach does not provide a quantitative estimate of the influence of each abiotic or biotic factor on the presence and the abundance of plant species, therefore making it challenging to obtain a numerical measure of its specific contribution to the observed ecological pattern. Furthermore, the application of this approach should always be guided by experts with solid ecological knowledge, as many of the associations identified may be statistically significant yet lack ecological meaning. Expert interpretation plays a crucial role in evaluating the validity and relevance of the outputs, ensuring that the resulting patterns are ecologically sound and contextually appropriate.  \\
In conclusion, this study contributes to a better understanding of plant species distribution and assemblage by offering useful insights and identifying meaningful association patterns among plant species and their abiotic and biotic drivers. The proposed approach proves effective in addressing the growing demand for automatic knowledge extraction from large, complex, and often noisy ecological datasets. Notably, its rapid processing capabilities enable the generation of a general overview of forest community structure and biodiversity patterns, highlighting dominant associations and potential ecological hubs.

Although challenges remain, particularly concerning the handling of ecological data complexity and the need for solid measures of rule significance and ecological relevance, the findings presented here represent a useful resource for future research exploring species interactions and community structure, especially in contexts where prior ecological knowledge is available. These results also underscore the importance of publicly accessible, high-quality ecological data repositories for advancing biodiversity research and conservation planning.

Future developments should explore the extension of the method toward association rules involving association of species as consequents, thereby shifting from single-species predictions to the identification of co-occurring species assemblages. This advancement would improve the ecological interpretability of the results and provide a more realistic representation of biodiversity dynamics and multi-species interactions within complex ecosystems.

\section*{\textbf{CRediT authorship contribution statement}}
\textbf{Valeria Aloisi:} Conceptualization, Investigation, Data curation, Methodology, Formal analysis, Software, Visualization, Writing - Original Draft, Writing - Review \& Editing. \textbf{Sergio Noce:} Data curation, Formal Analysis, Validation, Visualization, Writing - Original Draft, Writing - Review \& Editing. \textbf{Italo Epicoco:} Conceptualization, Investigation, Methodology, Writing - Review \& Editing, Supervision. \textbf{Cristina Cipriano:} Validation, Writing - Review \& Editing. \textbf{Massimo Cafaro:} Conceptualization, Methodology, Writing - Review \& Editing. \textbf{Giuseppe Brundu:} Conceptualization, Writing - Review \& Editing. \textbf{Lorenzo Arcidiaco:} Data curation, Writing - Review \& Editing. \textbf{Donatella Spano:} Conceptualization, Writing - Review \& Editing. \textbf{Giovanni Aloisio:} Conceptualization, Writing - Review \& Editing. \textbf{Simone Mereu:} Conceptualization, Validation, Writing - Review \& Editing.

\section*{\textbf{Declaration of competing interest}} The authors declare that they have no known competing financial interests or personal relationships that could have appeared to influence the work reported in this paper.

\section*{\textbf{Acknowledgments}}

The authors thank the Comando Unità Forestale, Ambientale e Agroalimentare (CUFAA) of the Arma dei Carabinieri for full access to the INFC 2015 data. 

\section*{\textbf{Funding}}

This research was partially funded under the National Recovery and Resilience Plan (NRRP), Mission 4 Component 2 Investment 1.4 – Call for tender No 3138 of 16 December 2021, rectified by Decree n.3175 of 18 December 2021 of Italian Ministry of University and Research funded by the European Union – NextGenerationEU. Project code CN\_00000033, Concession Decree No 1034 of 17 June 2022 adopted by the Italian Ministry of University and Research, CUP C83C22000550007, Project title “National Biodiversity Future Center – NBFC”.

\section*{\textbf{Data availability}}
INFC 2015 data: access upon consent by the Comando Unità Forestale, Ambientale e Agroalimentare (CUFAA) of the Arma dei Carabinieri.

WorldClim 2.1 database:
\url{https://geodata.ucdavis.edu/climate/worldclim/2_1/base/wc2.1_30s_bio.zip}

Version 3 of the Global Aridity index and potential Evapotranspiration database: \url{https://doi.org/10.6084/m9.figshare.7504448.v5}

Soil-related data: SoilGrids at \url{https://data.isric.org/geonetwork/srv/ita/catalog.search#/metadata/14e7c761-6f87-4f4c-9035-adb282439a44}

Lithology information: Italian lithological map provided by \cite{bucci2022new}.

NDVI data: derived using a  multi-year Sentinel-2 imagery available in the Google Earth Engine GEE archive \url{https://code.earthengine.google.com/}, \citep{gorelick2017google}.

Global high-resolution canopy height dataset: \url{langnico.github.io/globalcanopyheight/assets/tile_index.html}
 
Software repository: \url{https://github.com/CMCC-Foundation/ARM_for_Plants_Distribution}

\bibliographystyle{elsarticle-harv} 
\bibliography{Bibliography}

\appendix
\clearpage

\section{Supplementary Material}

\begin{longtable}{|l|l|}
\caption{Species and UNICODE list.} \\
\hline
\textbf{Species} & \textbf{UNICODE} \\
\hline
\endfirsthead

\hline
\textbf{Species} & \textbf{UNICODE} \\
\hline
\endhead
\hline
\endfoot

\hline
\endlastfoot 
Abies alba & ABIALB \\
Abies cephalonica & ABICEP \\
Abies sp. & ABISPE \\
Acacia sp. & ACASPE \\
Acer campestre & ACECAM \\
Acer gr. opalus & ACEOPA \\
Acer lobelii & ACELOB \\
Acer monspessulanum & ACEMON \\
Acer negundo & ACENEG \\
Acer platanoides & ACEPLA \\
Acer pseudoplatanus & ACEPSE \\
Aesculus hippocastanum & AESHIP \\
Ailanthus altissima & AILALT \\
Alnus cordata & ALNCOR \\
Alnus glutinosa & ALNGLU \\
Alnus incana & ALNINC \\
Alnus viridis & ALNVIR \\
Amelanchier ovalis & AMEOVA \\
Arbutus unedo & ARBUNE \\
Betula pendula & BETPEN \\
Betula pubescens & BETPUB \\
Betula sp. & BETSPE \\
Buddleja davidii & BUDDAV \\
Buxus sempervirens & BUXSEM \\
Buxus sp. & BUXSPE \\
Calicotome sp. & CALSPE \\
Carpinus betulus & CARBET \\
Carpinus orientalis & CARORI \\
Castanea sativa & CASSAT \\
Cedrus sp. & CEDSPE \\
Celtis sp. & CELAUS \\
Ceratonia siliqua & CERSIL \\
Cercis siliquastrum & CERSII \\
Chamaecyparis lawsoniana & CHALAW \\
Chamaerops humilis & CHAHUM \\
Clematis sp. & CLESPE \\
Cornus mas & CORMAS \\
Cornus sanguinea & CORSAN \\
Corylus sp. & CRYSPE \\
Cotinus coggygria & COTCOG \\
Cotoneaster sp. & COTSPE \\
Crataegus sp. & CRASPE \\
Cupressus arizonica & CUPARI \\
Cupressus sempervirens & CUPSEM \\
Cupressus sp. & CUPSPE \\
Elaeagnus angustifolia & ELAANG \\
Erica arborea & ERIARB \\
Erica scoparia & ERISCO \\
Erica sp. & ERISPE \\
Eucalyptus sp. & EUCSPE \\
Euonymus sp. & EUOSPE \\
Fagus sylvatica & FAGSYL \\
Ficus carica & FICCAR \\
Frangula sp. & FRASPE \\
Fraxinus excelsior & FRAEXC \\
Fraxinus ornus & FRAORN \\
Fraxinus oxycarpa & FRAOXY \\
Genista sp. & GENSPE \\
Hedera helix & HEDHEL \\
Hippophae rhamnoides & HIPRHA \\
Ilex aquifolium & ILEAQU \\
Juglans nigra & JUGNIG \\
Juglans regia & JUGREG \\
Juniperus communis & JUNCOM \\
Juniperus oxycedrus & JUNOXY \\
Juniperus phoenicea & JUNPHO \\
Laburnum sp. & LABSPE \\
Larix decidua & LARDEC \\
Laurus nobilis & LAUNOB \\
Ligustrum vulgare & LIGVUL \\
Lonicera sp. & LONSPE \\
Malus sp. & MALSPE \\
Mespilus germanica & MESGER \\
Morus sp. & MORSPE \\
Myrtus communis & MYRCOM \\
Nerium oleander & NEROLE \\
Olea europaea & OLEEUR \\
Opuntia ficus-indica & OPUFIC \\
Ostrya carpinifolia & OSTCAR \\
Paulownia tomentosa & PAUTOM \\
Phillyrea angustifolia & PHIANG \\
Phillyrea latifolia & PHILAT \\
Picea abies & PICABI \\
Picea sp. & PICSPE \\
Pinus laricio & PINLAR \\
Pinus brutia & PINBRU \\
Pinus cembra & PINCEM \\
Pinus excelsa (P.wallichiana) & PINEXC \\
Pinus halepensis & PINHAL \\
Pinus mugo & PINMUG \\
Pinus nigra & PINNIG \\
Pinus pinaster & PINPIN \\
Pinus pinea & PINPIE \\
Pinus radiata & PINRAD \\
Pinus sp. & PINSPE \\
Pinus strobus & PINSTR \\
Pinus sylvestris & PINSYL \\
Pinus uncinata & PINUNC \\
Pistacia lentiscus & PISLEN \\
Pistacia terebinthus & PISTER \\
Pittosporum tobira & PITTOB \\
Platanus hybrida & PLAHYB \\
Platanus orientalis & PLAORI \\
Populus alba & POPALB \\
Populus nigra & POPNIG \\
Populus sp. & POPSPE \\
Populus tremula & POPTRE \\
Populus xcanadensis & POPXCA \\
Prunus avium & PRUAVI \\
Prunus brigantiaca & PRUBRI \\
Prunus cerasifera & PRUCER \\
Prunus cocomilia & PRUCOC \\
Prunus laurocerasus & PRULAU \\
Prunus mahaleb & PRUMAH \\
Prunus padus & PRUPAD \\
Prunus serotina & PRUSER \\
Prunus sp. & PRUSPE \\
Prunus spinosa & PRUSPI \\
Pseudotsuga menziesii & PSEMEN \\
Pyracantha coccinea & PYRCOC \\
Pyrus sp. & PYRSPE \\
Quercus cerris & QUECER \\
Quercus crenata & QUECRE \\
Quercus frainetto & QUEFRA \\
Quercus ilex & QUEILE \\
Quercus petraea & QUEPET \\
Quercus pubescens & QUEPUB \\
Quercus robur & QUEROB \\
Quercus rubra & QUERUB \\
Quercus sp. & QUESPE \\
Quercus suber & QUESUB \\
Quercus trojana & QUETRO \\
Rhamnus alaternus & RHAALA \\
Rhamnus alpinus & RHAALP \\
Rhamnus catharticus & RHACAT \\
Rhamnus sp. & RHASPE \\
Rhus sp. & RHUSPE \\
Robinia pseudacacia & ROBPSE \\
Rosa sp. & ROSSPE \\
Rubia peregrina & RUBPER \\
Salix alba & SALALB \\
Salix caprea & SALCAP \\
Salix sp. & SALSPE \\
Sambucus nigra & SAMNIG \\
Sambucus racemosa & SAMRAC \\
Sorbus aria & SORARI \\
Sorbus aucuparia & SORAUC \\
Sorbus chamaemespilus & SORCHA \\
Sorbus domestica & SORDOM \\
Sorbus sp. & SORSPE \\
Sorbus torminalis & SORTOR \\
Spartium junceum & SPAJUN \\
Tamarix sp. & TAMSPE \\
Taxus baccata & TAXBAC \\
Thuja sp. & THUSPE \\
Tilia cordata & TILCOR \\
Tilia platyphyllos & TILPLA \\
Tilia sp. & TILSPE \\
Ulmus glabra & ULMGLA \\
Ulmus minor & ULMMIN \\
Ulmus sp. & ULMSPE \\
Viburnum lantana & VIBLAN \\
Viburnum tinus & VIBTIN \\
Vitis sp. & VITSPE \\
\hline
\end{longtable}
\begin{table}[ht]
\centering
\caption{Lithological classes: codes and descriptions. From \cite{bucci2022new}}
\label{tab:lithology}
\begin{tabularx}{\textwidth}{c c X}
\toprule
\textbf{ID} & \textbf{Code} & \textbf{Description (English)} \\
\midrule
1  & Ad  & Anthropogenic deposits \\
2  & CM  & Chaotic – mélange \\
3  & Al  & Alluvial, lacustrine, swamp and marine deposits. Eluvial and colluvial deposits \\
4  & Gd  & Glacial drift \\
5  & E   & Evaporite \\
6  & Ssr & Siliciclastic sedimentary rocks \\
7  & Mw  & Mass wasting material \\
8  & Li  & Lakes and Ice \\
9  & Lb  & Lavas and basalts \\
10 & M   & Marlstone \\
11 & Pr  & Pyroclastic rocks \\
12 & Cr  & Carbonate rocks \\
13 & Ir  & Intrusive rocks \\
14 & Nsr & Non-schistose metamorphic rocks \\
15 & Sr  & Schistose metamorphic rocks \\
16 & Ccr & Consolidated clastic rocks \\
17 & Ucr & Unconsolidated clastic rock \\
18 & SM  & Mixed sedimentary rocks \\
20 & B   & Beaches and coastal deposits \\
\bottomrule
\end{tabularx}
\end{table}

\begin{sidewaystable}    
\caption{Database description and data sources.}
\label{TableDB}
\centering
\begin{adjustbox}{max width=\textheight, max totalheight=\textwidth, center}
\begin{tabular}{|c|c|c|c|c|c|}
\hline
\textbf{Category} & \textbf{Code} & \textbf{Description} & \textbf{Units} & \textbf{Source} & \textbf{Spatial Resolution} \\ 
\hline
\multirow{6}{*}{Info Plot} 
& idplot & ID PLOT & - & - & - \\ \cline{2-6}
& P\_UNICODE & Species Code & - & - & - \\ \cline{2-6}
& P\_BAARSP & Species Basal Area & m²/ha & INFC 2015 & - \\ \cline{2-6}
& P\_BASRPL & Plot total Basal Area & " & " & - \\ \cline{2-6}
& P\_LATITU & Latitude WGS84* & Decimal Degree & " & - \\ \cline{2-6}
& P\_LONGIT & Longitude WGS84* & " & " & - \\ 
\hline
\multirow{3}{*}{Soil} 
& S\_SGWC33 & Volumetric Water Content t -33 kPa 5-15 cm & $10^{-3}$ cm\textsuperscript{3}/cm\textsuperscript{3} & SoilGrids & 250 m \\ \cline{2-6}
& S\_SGPHWA & pH Water 5-15 cm & - & " & " \\ \cline{2-6}
& S\_ITLITO & Lithology Class & - & Lithological map of Italy\citep{bucci2022new} & 1:100000 \\ 
\hline
\multirow{20}{*}{Climate} 
& C\_WC0001 & Annual Mean Temperature (Bio1) & °C & WorldClim 2.1 & 30" (approx 1 km) \\ \cline{2-6}
& C\_WC0002 & Mean Diurnal Range (Bio2)  & " & " & " \\ \cline{2-6}
& C\_WC0003 & Isothermality (Bio3) & \% & " & " \\ \cline{2-6}
& C\_WC0004 & Temperature Seasonality (Bio4) & °C & " & " \\ \cline{2-6}
& C\_WC0005 & Max. Temperature of Warmest Month (Bio5) & " & " & " \\ \cline{2-6}
& C\_WC0006 & Min. Temperature of Coldest Month (Bio6) & " & " & " \\ \cline{2-6}
& C\_WC0007 & Temperature Annual Range (Bio7) & " & " & " \\ \cline{2-6}
& C\_WC0008 & Mean Temperature of Wettest Quarter (Bio8) & " & " & " \\ \cline{2-6}
& C\_WC0009 & Mean Temperature of Driest Quarter (Bio9) & " & " & " \\ \cline{2-6}
& C\_WC0010 & Mean Temperature of Warmest Quarter (Bio10) & " & " & " \\ \cline{2-6}
& C\_WC0011 & Mean Temperature of Coldest Quarter (Bio11) & " & " & " \\ \cline{2-6}
& C\_WC0012 & Annual Precipitation (Bio12) & mm & " & " \\ \cline{2-6}
& C\_WC0013 & Precipitation of Wettest Month (Bio13) & " & " & " \\ \cline{2-6}
& C\_WC0014 & Precipitation of Driest Month (Bio14) & " & " & " \\ \cline{2-6}
& C\_WC0015 & Precipitation Seasonality (Bio15) & \% & " & " \\ \cline{2-6}
& C\_WC0016 & Precipitation of Wettest Quarter (Bio16) & mm & " & " \\ \cline{2-6}
& C\_WC0017 & Precipitation of Driest Quarter (Bio17) & " & " & " \\ \cline{2-6}
& C\_WC0018 & Precipitation of Warmest Quarter (Bio18) & " & " & " \\ \cline{2-6}
& C\_WC0019 & Precipitation of Coldest Quarter (Bio19) & " & " & " \\ \cline{2-6}
& C\_ARIIND & Aridity Index & - & Global Aridity Index and Pot. Evap. Database\citep{zomer2022version} & " \\ 
\hline
\multirow{3}{*}{Earth Observation} 
& E\_CANHEI & Canopy Height & m & ETH Global Sentinel-2 Canopy Height & 10 m \\ \cline{2-6}
& E\_MDVIME & Maximum NDVI 90 Percentile 2017-2024 & - & Sentinel 2 & " \\ \cline{2-6}
& E\_SDVIME & Standar Deviation NDVI Max 2017-2024 & - & " & " \\ 
\hline
\end{tabular}
\end{adjustbox}
\footnotesize
\textsuperscript{* Latitude and Longitude data cannot be provided}
\end{sidewaystable}

\end{document}